\documentclass{article}

\usepackage{graphicx}
\usepackage{SIunits}
\usepackage{epstopdf}

\usepackage[T1]{fontenc}       
\usepackage{lmodern}
\usepackage[ margin=2cm]{geometry}

\usepackage{color}
\newcommand{\ch}[1]{{ {#1} }}

\author{A. W. Schell\textsuperscript{1*}, M. Svedendahl\textsuperscript{1*}, R. Quidant\textsuperscript{1,2} \\ \vspace{-1mm} \small 1. ICFO-Institut de Ciencies Fotoniques, \\ \vspace{-1mm} \small Barcelona Institute of Science and Technology\\ \small 08860 Castelldefels (Barcelona), Spain \\  \vspace{-1mm} \small 2. ICREA-Instituci\`{o} Catalana de Recerca i Estudis Avan\c{c}ats \\ \vspace{-1mm} \small 08010 Barcelona, Spain\\ \small * These authors contributed equally \\ \small Email: andreas.schell@icfo.eu \vspace{-10mm} \normalsize}

\title{Quantum Emitters in Hexagonal Boron Nitride Have Spectrally Tunable Quantum Efficiency}

\date{~}
\begin{document}
\maketitle

\begin{abstract}
\large
Understanding the properties of novel solid-state quantum emitters is pivotal for a variety of applications in field ranging from quantum optics to biology. Recently discovered defects in hexagonal boron nitride are especially interesting, as they offer much desired characteristics such as narrow emission lines and photostability. Here, we study the dependence of the emission on the excitation wavelength. We find that, in order to achieve bright single photon emission with high quantum efficiency, the excitation wavelength has to be matched to the emitter. This is a strong indication that the emitters possess a complex level scheme and cannot be described by a simple two or three level system. Using this excitation dependence of the emission, we thus gain further insight to the internal level scheme and demonstrate how to distinguish different emitters both spatially as well as in terms of their photon correlations.
\end{abstract}
\large
Research on quantum emitters in solid-state materials has gained 
momentum with the discovery of a variety of new emitters  in recent years  and the first successful attempts to 
engineer their properties.~\cite{Aharonovich2016}
Such solid-state emitters 
are a promising alternative 
to trapped atoms and ions in quantum information processing~\cite{Monroe2002} as they feature narrow 
transition lines and long coherence times while having advantages in many aspects, in particular 
in terms of scalability and miniaturization.~\cite{Obrien2009}
As emitters in a solid material can easily be moved around and brought into the vicinity of 
other structures, they resemble nearly ideal probes and can be used in 
various sensing experiments, e.g., for measuring their electric and magnetic 
environment.~\cite{Balasubramanian2008,Taylor2008,Dolde2011,Schell2014}
Furthermore, solid-state quantum emitters can be extremely photostable and 
can serve as an alternative to organic dyes as biomarkers.~\cite{Fu2007}

Recently, quantum emitters hosted in atomically thin, so 
called two-dimensional, materials have been 
discovered.~\cite{Chakraborty2015,Srivastava2015,He2015,Koperski2015,Tonndorf2015,Tran2015}
These materials are promising for a variety of applications, ranging 
from optical switching to biomedical applications.~\cite{Tsai2013,Chen2015}
One of these emitters, found in hexagonal boron nitride (hBN), has turned out to be
a bright, photostable, room temperature single photon source. Furthermore, it posses 
narrow emission lines and can be excited using non-linear processes.~\cite{Tran2015,Schell2016,Sontheimer2017}
Thanks to all these attractive properties, first attempts to integrate such defects in 
photonic structures were successfully carried out.~\cite{Schell2017}
Nevertheless, up to date the details of the emitters'  level structure remain elusive. 
So far, single photon emitters in hBN have been studied at different 
temperatures~\cite{Tran2015,Sontheimer2017,Kianinia2017} and
different emission wavelengths~\cite{Bourrellier2016,Tran2016}  
\ch{and mechanisms to alter the emission 
wavelength optically has been investigated.~\cite{Shotan2016}}
In addition, first principle calculations using \ch{group and} density-functional theory 
of the energy levels have been carried out~\cite{Li2017,Tawfik2017,Abdi2017} to get insight 
into the atomic structure of the defects. 
A study of the polarization selection rules of the zero phonon line of the defects revealed
a misalignment of emission and absorption dipole -- a strong indication of a multi-level system.\cite{Jungwirth2017}
\ch{In order to control the creation of these defects and to understand their atomic origin, 
defect formation has been studied for example at different annealing temperatures,~\cite{Tran2016} and different atmospheres,~\cite{Tran2016} 
and with etching and ion implantation.~\cite{Chejanovsky2016} 
Despite these efforts, the atomic origin remains unknown and more information 
ont he properties of the emitters needs to be gathered.}

\begin{figure}
{\includegraphics{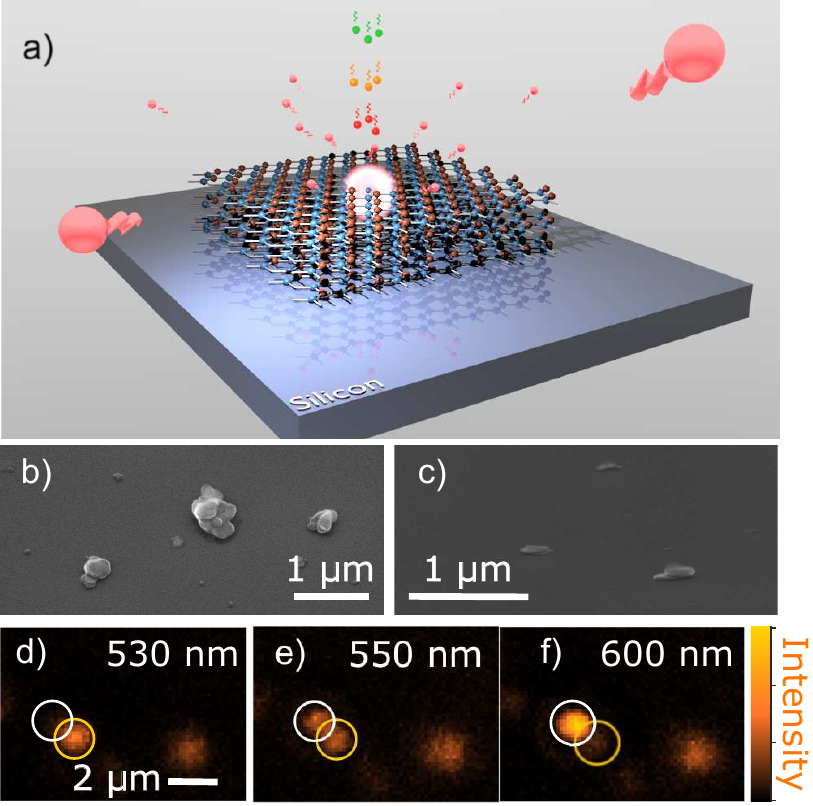}}
\centering
\caption{\label{fig1} (a) Artistic illustration of the experiment. The emitters are excited by different wavelengths and the photoluminescence is collected and studied in a confocal microscope.  
(b,c) Scanning electron micrographs 
of different hexagonal boron nitride nanoflakes at (b) 35$^\circ$ and (c) 80$^\circ$  incidence. 
(d-f) 
Confocal scans of a section of the sample, using \unit{530}{\nano\meter}, \unit{550}{\nano\meter}, 
and \unit{600}{\nano\meter} excitation, respectively. White (left) and yellow (right)
circles indicate two different single photon emitters. }
\end{figure}

Here, we study the level structure of \ch{defects in multilayer} hBN \ch{flakes} by photoluminescence excitation (PLE) 
spectroscopy on single emitters. By varying the excitation wavelength while monitoring emission 
intensity and emission spectrum (Figure \ref{fig1}a), we gain knowledge on the 
level structure of the emitter and demonstrate the spectral dependence of the quantum efficiency.

The optical setup consists of a home-built confocal microscope, fiber coupled 
to an optical parametric oscillator (OPO, Opium, Radiantis) in another lab, a spectrometer (Andor Shamrock 303i)
with a cooled CCD camera (Andor iDus), 
and two avalanche photodiodes (APDs, Micro Photon Devices) in a Hanbury Brown and Twiss (HBT) configuration. 
The OPO emission was used to excite the hBN nanoflakes through a 0.85 NA 60X magnification objective 
(Edmund Optics).  The laser powers are measured 
in front of the slightly overfilled objective lens. The emitted photoluminescence was collected through the same 
obective and separated from the excitation light by means of a 50:50 beamsplitter. 
The photoluminescence was filtered by a \unit{633}{\nano\meter} longpass filter 
and fiber coupled either to the spectrometer or the APDs. The samples was mounted on a x-y-z piezo-translation 
stage to facilitate confocal photoluminescence scans.

The samples used were hBN nanoflakes (in ethanol/water, Graphene Supermarket) 
drop casted on clean silicon wafers. After evaporation of the solvent, 
the samples were annealed in a nitrogen environment at \unit{500}{\degree\Celsius} for four hours
\ch{in an oven (Unitemp RTP-150)}. 
The annealing increase the number of emitters on the substrates, although bright 
single photon emitters were already found on non-annealed samples. 
The scanning electron microscope (SEM) images (Figure \ref{fig1}b-c) show typical nanoflakes 
around \unit{100-500}{\nano\meter} in size, with thicknesses estimated 
around \unit{10-40}{\nano\meter}. 
Additional TEM and Raman characterisation of the nanoflakes (available in the Supporting Information) confirms the composition and crystallinity of the nanoflakes. 

Panels (d-f) of Figure \ref{fig1} illustrate the effect that is investigated in more detail in the following. 
Confocal images show two different, adjacent, quantum emitters 
located by the white and yellow circles. Specifically, the brightest emitter in panel (d), 
corresponding to \unit{530}{\nano\meter} excitation laser wavelength, is only weakly visible in 
the scans using \unit{600}{\nano\meter} illumination. Conversely, the strongest emitter in the \unit{600}{\nano\meter} case is not visible in the \unit{530}{\nano\meter} scans. 
However, both these emitters appear in the confocal scans using \unit{550}{\nano\meter} light. 
These strikingly different images imply that they have very different excitation spectra and, therefore, level structure.

\begin{figure}
{\includegraphics{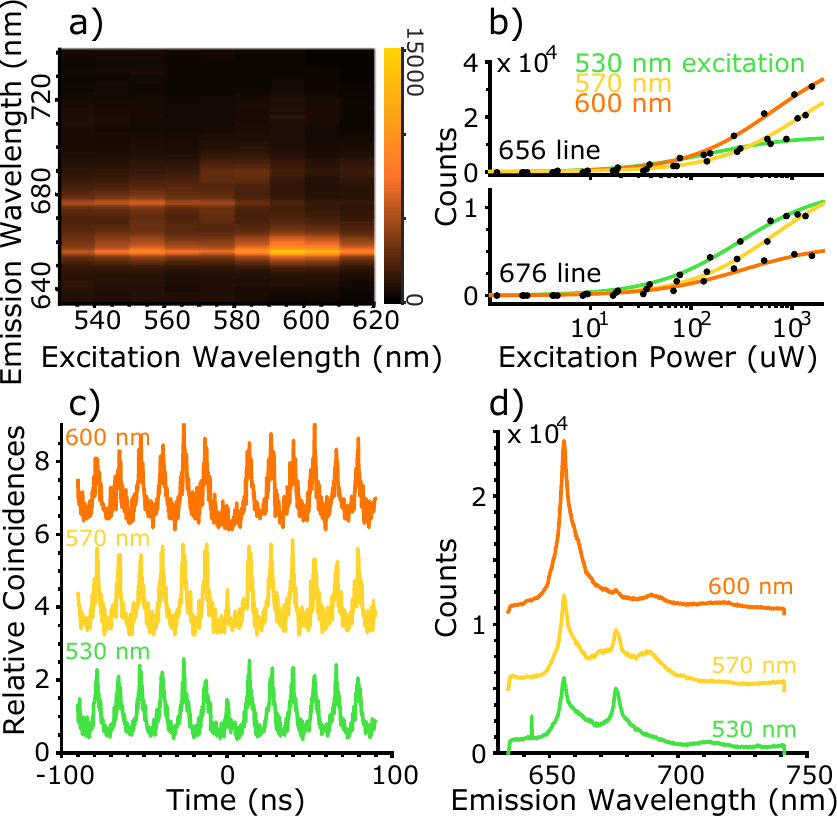}}
\centering
\caption{\label{fig2}  (a) Photoluminescence from a hBN nanoflake for excitation wavelengths ranging from \unit{530}{\nano\meter} to \unit{620}{\nano\meter}. 
Two distinct peaks are visible.
(b) Saturation curves for the two main emission lines, located at \unit{656}{\nano\meter} and \unit{676}{\nano\meter}, using 
\unit{530}{\nano\meter}, \unit{570}{\nano\meter}, and \unit{600}{\nano\meter} excitation, respectively. 
The solid lines are fits to the data, black dot data points.
Interestingly, the saturation intensity as well as the maximum achievable count rate change with 
excitation wavelength.
(c) Antibunching measured for \unit{530}{\nano\meter}, \unit{570}{\nano\meter} and \unit{600}{\nano\meter} 
excitation, respectively. The depth of the antibunching changes with wavelength.
(d) Emission spectra corresponding to the measurements in (c).}
\end{figure}

To investigate this effect in more detail we studied the photoluminescence of several 
nanoflakes with emission lines ranging from \unit{630}{\nano\meter} to \unit{740}{\nano\meter},
 using \unit{530}{\nano\meter}-\unit{620}{\nano\meter} excitation 
from the OPO. An overview of the photolumiscence spectra of a nanoflake as a function of 
excitation wavelength can be seen in Figure \ref{fig2}a. 
In this specific case, two emission lines stand out: at \unit{656}{\nano\meter} and 
\unit{676}{\nano\meter}, although the spectra include more features. As in Figure \ref{fig1},
 these two emission lines vary in strength depending on the excitation wavelength. Specifically, 
the line at \unit{656}{\nano\meter} is the brightest under \unit{590}{\nano\meter} 
illumination, while the \unit{676}{\nano\meter} line is most efficiently excited at \unit{540}{\nano\meter}. 

Furthermore, we studied the power dependence of the emission using different excitation wavelengths. 
This is important, as the count rate for an emitter driven in saturation is directly linked to the quantum efficiency. 
Much like the excitation spectra, the saturation curves (Figure \ref{fig2}b) 
also show a strong dependence on the excitation wavelength. We fitted the detected counts at the spectral maxima 
of the two main emission peaks to the expected behavior of the count rate:

\begin{equation}
\label{eq1}
C_{out} = R_{\infty}\frac {I_{in}}{I_{in}+I_{s}}+B~, 
\end{equation}

where $R_{\infty}$ is the maximum rate out, $I_{in}$ is the input power, $I_s$ is the 
saturation power and $B$ is the constant background. For the \unit{656}{\nano\meter}  (\unit{676}{\nano\meter})  line the 
fits yield $R_{\infty}$ of 13k (12k), 40k (14k) and 44k (0.6k) counts for 
\unit{530}{\nano\meter}, \unit{570}{\nano\meter}, and \unit{600}{\nano\meter}
excitation, respectively.  Furthermore, the saturation powers are 
\unit{140}{\micro\watt} (\unit{520}{\micro\watt}),  \unit{1250}{\micro\watt} (\unit{650}{\micro\watt}), 
and \unit{650}{\micro\watt} (\unit{320}{\micro\watt}), respectively.
As in a two level system, the value $R_{\infty}$  is only governed by the decay process, 
the different $R_{\infty}$ indicate that this system cannot be treated as a simple two level system.

However, with these measurements it is not clear whether the two lines in the emitted spectrum stem 
from the same single emitter or two nearby emitters that could not, in contrast to the 
emitters shown in Figure~\ref{fig1}, be spatially resolved.
In order to discriminate between these two possibilities, we performed antibunching 
measurements at \unit{300}{\micro\watt} excitation power (Figure \ref{fig2}c) using the 
HBT setup. Comparing the results with the photoluminsescence spectra in Figure \ref{fig2}d, 
we can conclude that when a single line is dominating (at \unit{600}{\nano\meter} excitation), 
the antibunching signature is clear with a peak missing at $\Delta$t=0 in the coincidence graphs. 
Conversely, as on changing the wavelength, the \unit{676}{\nano\meter} line grows in strength relative to the 
\unit{656}{\nano\meter} line, a peak in the correlation function at $\Delta$t=0 appears and increases in height. 
The results therefore imply that the two lines in this case come 
from two different emitters located within the same diffraction limited spot, possibly within the same nanoflake. 

\begin{figure}
\centering
{\includegraphics{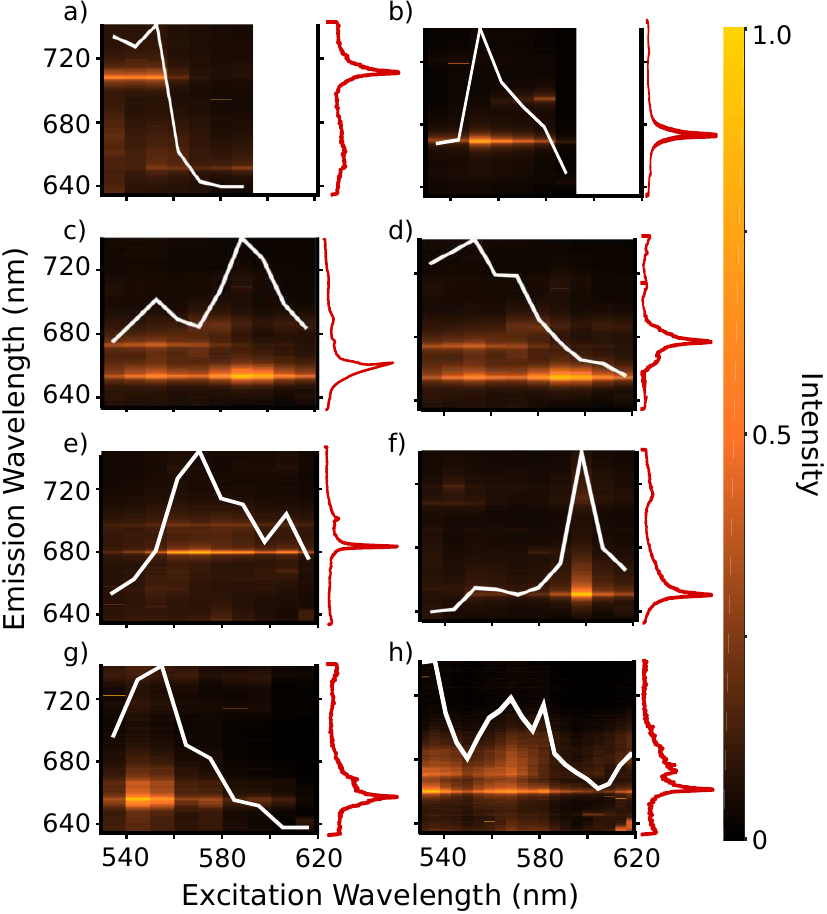}}
\caption{\label{fig3} 
\ch{Panels a-h show the excitation dependence of the emission for eight emission lines. 
The white line indicates the brightness of the line for the different excitation wavelengths while in 
the background the full spectra are shown as a color map.
Again, a strong dependence of the photoluminescence intensity on excitation wavelength is visible, while the photoluminsecence spectrum did not change significantly.
The emission spectra shown are the normalized measurements at the most efficient excitation wavelength.
Note that the lines in panels c and d were acquired at the same spot. (See Supporting Information 
for more spectra)}}
\end{figure}

Figure \ref{fig3} shows the excitation and emission spectra for several hBN nanoflake 
single photon emitters. Clearly, different emitters have different emission 
as well as excitation spectra. 
No variation of the width or location of the emission lines was found as a function 
of the excitation wavelength. Rather, these emitters have stable and reproducible 
emission during the scans with narrow lines at a well-defined wavelength. 

\begin{figure}
{\includegraphics{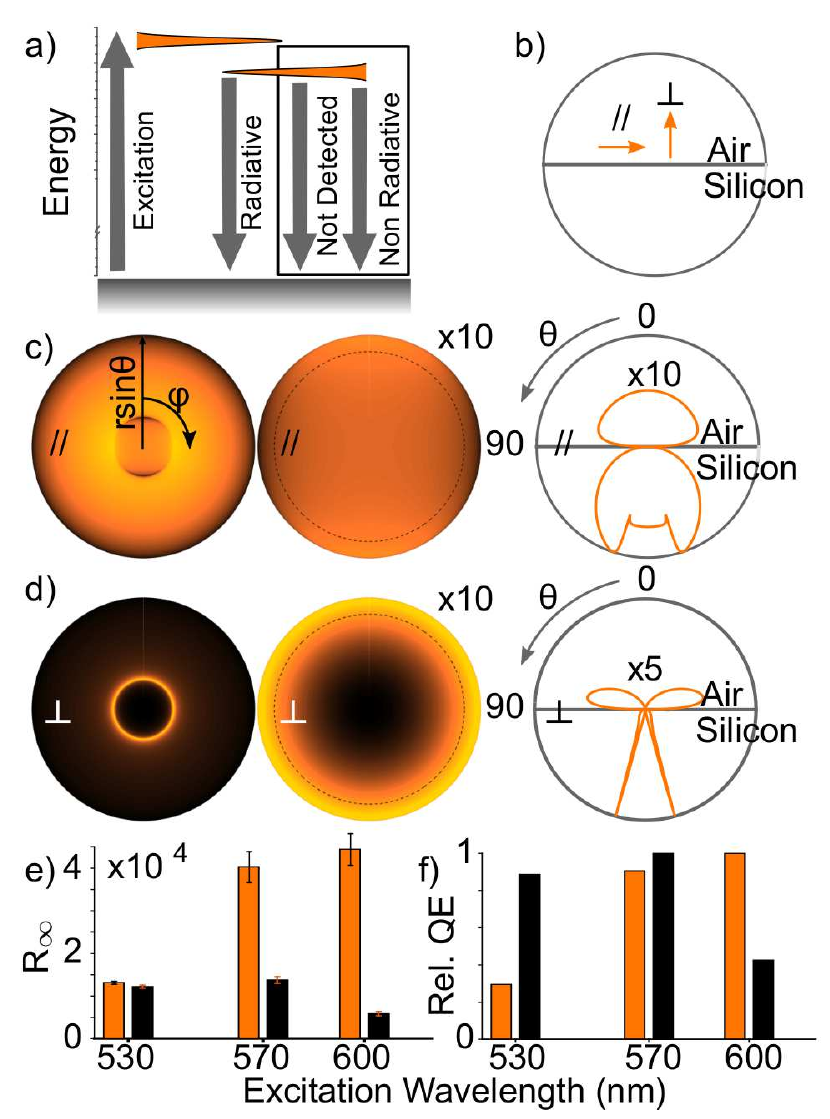}}
\centering
\caption{\label{figx}(a) Basic level scheme for an emission process. Upon excitation,
the emitter decays non radiatively or emits a photon, which can either be 
detected or not detected.
(b) Emission pattern of a dipole near an air/silicon interface. The arrows indicate the different orientations of the dipole, parallel or perpendicular to the interface. 
(c,d) Two dimensional plots of the emission patterns into silicon (left) and air (center). $r \sin{\theta}$ runs from 0 to 1 while 
$\phi$ runs from 0 to $2\pi$. (right) The $\theta$-dependence of the emission. 
(e) Saturation count rate $R_{\infty}$ for the 656 nm (orange, left side) and 676 nm (black, right side) line studied in Figure~\ref{fig2}. Error bars are from the fit to Equation~(\ref{eq1}).
(f) Corresponding relative quantum efficiency.
}
\end{figure}

The excitation spectra typically contain a distinct resonance, indicating that the 
excitation brings the system into an intermediate energy level, located in the band gap of hBN. 
From this level, the excitation is transferred to the final excited state, 
from which photoluminescence occurs. There is also the possibility, that the excitation 
decays to the ground without emitting a photon at all, or that the photon emitted is not 
detected because it does not fall in the solid angle the microscope objective collects or 
it falls out of the detection range (see Figure~\ref{figx}a).

To assess the absolute quantum efficiency of the emitters we compared the 
detected photons with the number of photons an ideal emitter would provide. For this, 
we assume that the ideal emitter would provide one photon per cycle, that is, 80 million counts in 
our case (here, we neglected re-excitation in the same pulses, as our pulses are much 
shorter than the excited state's lifetime). By characterizing the efficiency of our microscope, and taking into account the radiation patterns of dipolar emitters near a silicon/air interface, we deduce the quantum efficiency of the emitters. The emission patterns in Figure~\ref{figx}c-d are calculated for an air/silicon interface, at \unit{660}{nm} with $n_{Si}=3.8$ and neglecting the losses in the silicon.~\cite{Lieb2004} 
For the emitters in Figure~\ref{fig2}, the quantum efficiency is estimated to be 0.5-1.0 and 0.2-0.6 for the \unit{656}{nm} and \unit{676}{nm} lines, respectively, for the most efficient excitation wavelengths. Here, we assumed an in-plane dipole (see Figure~\ref{figx}b-d) no  further from the silicon surface than \unit{40}{nm}. 

The relative spectral dependence of the quantum efficiencies for these emitters is shown in Figure~\ref{figx}f \ch{(for more data, please see Supporting Information)}. The large variation observed here clearly shows that when working with single defects in hBN not only the emission should be considered, but also the excitation wavelength is of importance. Here, we emphesize that this is fundamentally different from a change in excitation  efficiency as can be observed in many single photon emitters, for example nitrogen vacancy centers in diamond.~\cite{Davies285} 
We find that the quantum efficiency has a different spectral dependence compared to the  relative excitation efficiency.
Naturally, a wavelength-dependent absorption profile will affect the emitted power from a given quantum emitter, but once excited the quantum efficiency from this level is given by the different decay channels. The quantum efficiency is therefore not expected to vary with the excitation wavelength for a simple two-level system. Our results thus imply that there are multiple levels that compete with each other for the excitation energy and that this competition varies in strength over the studied spectral range, ultimately affecting the quantum efficiency of the emission. 

\begin{figure}
{\includegraphics{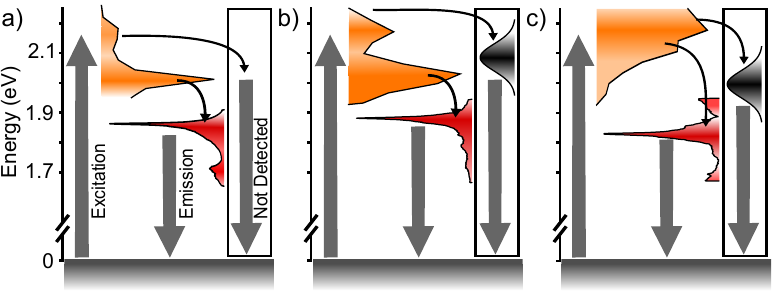}}
\centering
\caption{\label{fig4} Possible energy diagrams. 
(a) A simple energy diagram that only considers the excitation and emission 
spectra in Figure \ref{fig3}b. 
(b-c) More complex energy diagrams that also considers the various $R_{\infty}$ 
found from the data in Figure \ref{fig2}b. The emitters are excited to the 
intermediate level (left spectrum) and decay either into the emission level 
(center) or into another channel (right Gaussian spectrum), from which we do 
not detect any photons, either because the photon wavelength is outside our detection window
or because the decay is non-radiative.}
\end{figure}

From these findings, we can now try to get further insight on the level system of the defects.
Figure \ref{fig4}a show a schematic of possible internal energy levels that includes radiative and 
non-radiative decay channels and correctly accounts for the wavelength dependence of the 
excitation efficiency.
However, due to the varying wavelength dependence of $R_{\infty}$ and $I_{sat}$  presented 
in Figure~\ref{fig2}b, it is more likely that there are more (dark) 
levels involved in the energy dissipation as shown in Figure~\ref{fig4}b-c. 
These levels could possibly be excited directly by the laser, and compete with the 
intermediate level for direct excitation, or be excited via the intermediate level, 
and compete with the radiative decay channel. Depending on the position of this additional level, 
one ends up with the different excitation spectra and $R_{\infty}$, as for the emitters in Figure~\ref{fig2}. 

In conclusion, we have shown that the excitation efficiency as well as quantum efficiency of 
emitters in hBN are strongly wavelength dependent. This can be used to separate closely spaced 
emitters as shown in Figure~\ref{fig1}d-f and Figure~\ref{fig2}, which together with their photostability makes these emitters a potential candidate for super-resolution imaging techniques.~\cite{huang2009super} Another consequence, for experiments which aim to use defects in hBN as efficient emitters, 
is that the excitation wavelength has to be tuned to gain the highest quantum efficiency. This is especially important for quantum information processing techniques. 

Our findings suggest that the level structure of the defects in hBN is much  more
complex than a two or three level system. More details on the  level structure could be probed using multiple wavelength excitation and pump-probe approaches.

\section*{Acknowledgement}

The authors acknowledge financial support from the European Community's Seventh Framework Program under grant QnanoMECA (64790), Fundaci\'{o} Privada Cellex, and the Spanish Ministry of Economy and Competitiveness, through the ``Severo Ochoa'' Programme for Centres of Excellence in R\&D (SEV-2015-0522) and grant FIS2016-80293-R, and Swedish Research Council (637-2014-6894). Dr. Fei Ye and Dr. Stephan Steinhauer are respectively acknowledged for their assistance with TEM operation and analysis.

\clearpage

\renewcommand{\thefigure}{S\arabic{figure}}
\setcounter{figure}{0}

\begin{center}
{\bfseries \LARGE 
Quantum Emitters in Hexagonal Boron Nitride 
Have Spectrally Tunable Quantum Efficiency: 
Supporting Information}
\end{center}

\section*{TEM and Raman characterization}

The transmission electron microscope images where collected using a Jeol JEM 2100F microscope. The hBN nanoflakes were drop casted on a TEM grid (copper with carbon/formvar, 200 mesh, TedPella) and left to evaporate overnight. The diffraction pattern was processed using the CrysTBox software,~\cite{Klinger2015} which matched the collected pattern with reported values of the hBN diffraction lines. We note that the (100) and (110) lines are typically the orientations found in these systems.~\cite{Tran2015s, Guo2012}

The Raman spectra were collected using an In Via Raman microscope (Renishaw), using an 100X objective to focus a \unit{532}{\nano\meter}, \unit{12}{\milli\watt}, optical beam. Various sizes of hBN nanoflake ensembles were interrogated and all showed the typical E$_{2g}$  \unit{1367}{\centi\meter}$^{-1}$ line, with nanoflakes with fewer layers yielding a slightly broader peak.~\cite{Tran2015, Nemanich1981}

\begin{figure}
{\includegraphics{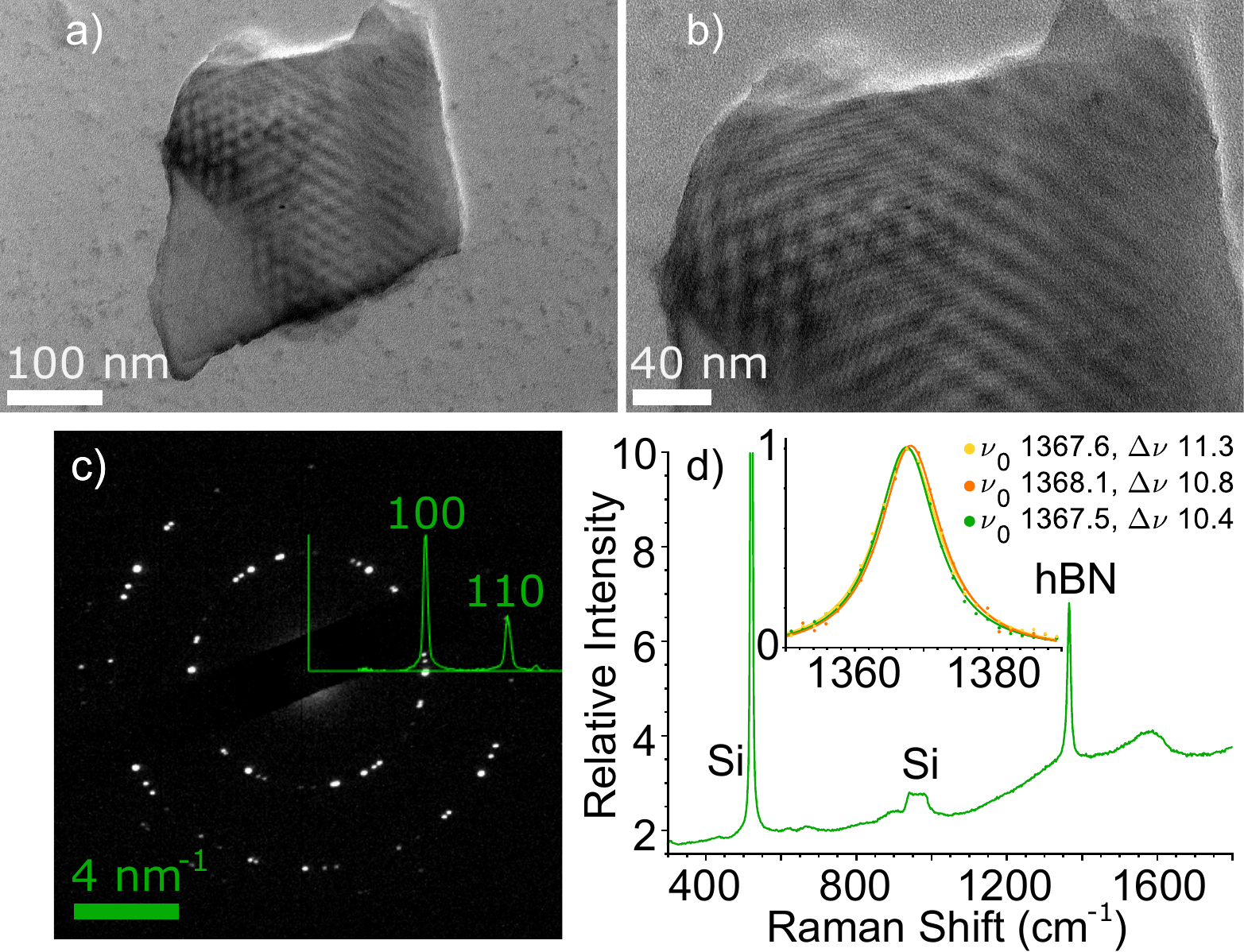}}
\centering
\caption{\label{fig2s} (a-b) Transmission electron microscope images the of hBN nanoflakes. (c) The corresponding diffraction pattern from the area in (b), after background correction and pattern recognition using CrysTBox.\cite{Klinger2015} The pattern is not expected to show monocrystallinity, as there are several nanoflakes in the interrogation area, but it clearly shows the expected (100) and (110) diffraction lines that fit with reported data for hBN. (d) Raman spectra of hBN nanoflake ensembles on a silicon wafer. The data show the E$_{2g}$ vibrational mode near \unit{1367}{\centi\meter}$^{-1}$, together with the expected Si lines. The inset show normalized spectra of the hBN line together with two measurements of ensembles of nanoflakes with fewer layers (yellow and orange), which show slightly broader peaks. The data points are experimental data, while the solid lines are Lorentzian fits, with peak position $\nu_0$ and full width at half maximum $\Delta\nu$.
}
\end{figure}

\begin{figure}
{\includegraphics{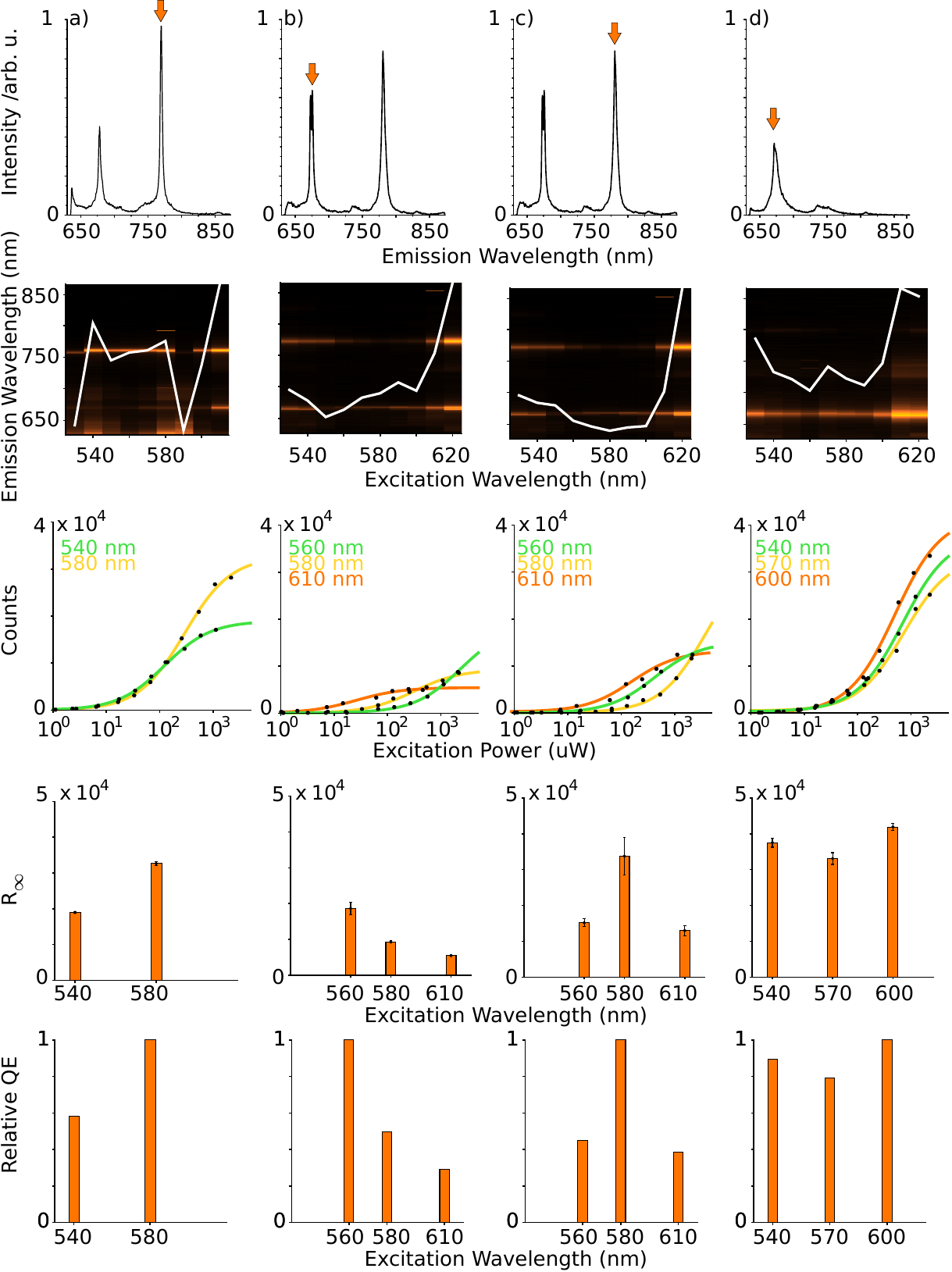}}
\centering
\caption{\label{fig1s} (a-d) show the dependence of the emission on the excitation wavelength for different emitters in hexagonal boron nitride. 
In the top row, the emission spectra at the most efficient wavelength are shown. Arrows indicate the lines analysed in the following.
In the second row, the efficiency of the excitation is shown in white and the back the full spectra re shown in a color map
in analogy to Figure 3 of the main text.
The third row shows the excitation power dependence of the emission under excitation with selected wavelengths (see Figure 2 b in the main text).
The last two rows show the saturation count rate $R_{\inf}$ extracted from these data and the relative quantum efficiency (see Figure 4 e,f of the main text).
}
\end{figure}

\end{document}